\documentclass[11pt,twoside]{article}

\usepackage{asp2006}
\usepackage{epsf}
\usepackage{graphics}
\usepackage{lscape}

\def\edcomment#1{\iffalse\marginpar{\raggedright\sl#1\/}\else\relax\fi}
\reversemarginpar

\begin{document}
\title{Statistical properties of flares and sunspots over the solar cycle}
\author{M. Temmer}
\affil{Kanzelh\"ohe Observatory/IGAM, Institute of Physics, University of Graz, A-8010 Graz, Austria}

\begin{abstract}
The present paper reviews results derived from statistical studies on solar activity indices. The prolonged minimum phase of cycle 23 raised the question of peculiarities inherent in cycle 23. The most important solar activity index is the relative sunspot number and though most of other indices are closely related, shifts are obtained between their peak activity of the order of 1--2 years. These shifts reveal a 22-yr pattern which can be attributed to solar interior or dynamo related processes. The minimum phase of cycle 23 is not found to be exceptional. Investigating the relative sunspot numbers over the past 150 years, solar cycles of more prolonged minima are observed. Since 1920 solar activity is quite high (``modern maximum'') and cycle 23 might be the herald of the end of this phase.
\end{abstract}

\vspace{-0.5cm}
\section{Introduction}

Solar cycle 23 shows characteristics of a prolonged minimum period. The rising activity of cycle 24 is observed to be late, although first spots of opposite polarity with respect to cycle 23 have been already observed in 2007. Still, activity remains, with exception of some single active regions, quite weak. [At the time of writing, beginning of 2010, the solar activity is regaining strength and several C- and M-class flares were already observed.] Can we use statistics to draw inferences about the characteristics of the solar activity cycle and its present behavior - is this solar minimum peculiar?

Space age started in the early 60s, since then sun-dedicated missions provide us with new data beyond restriction due to the limited spectral window of the Earth's atmosphere, most important for X-ray and EUV observations of solar flares, and white light coronagraph observations of coronal mass ejections. Such activity indices are collected since the recent past, whereas from ground-based observations sunspot numbers are recorded since 1600, sunspot areas since 1874 (Greenwich Observatory), and H$\alpha$ flares since 1940s.

Compared to the activity measurements based on solar flares and coronal mass ejections (CMEs), the physical relevance of relative sunspot numbers is often doubted especially the complexity of active regions might not be adequately taken into account. To compensate for limitations such as interpretation of the observer, seeing conditions, etc., the Solar Influences Data Analysis Center (SIDC; Vanlommel et al. 2004) is responsible for the daily international Sunspot Number which is computed as a weighted average of measurements made from a network of more than 25 cooperating observatories (Clette et al. 2007). Today, much more sophisticated measurements of solar activity are made routinely, but none has the link with the past that sunspot numbers have.

For the recognition of patterns, randomness and uncertainties in these long-term data sets statistical tools are applied. The results can then be used for prediction and forecasting on the basis of models. This information also provides valuable constraints for solar dynamo theories.

\section{Solar activity based on Sunspot Number}

The temporal behavior of the solar cycle is defined on the basis of the number of sunspots and sunspot groups  observed on the visible solar disk, indicated by the relative sunspot numbers $R$. The remarkable relevance of $R$ lies in particular in the fact that it represents the longest time series of solar activity indices, which is derived directly from observations of the Sun (in contrast to proxy data like cosmogenic isotopes). Thus, relative sunspot numbers provide the foundation of a continuous data set for research on the solar cycle and its long-term variations. $R$ is defined by
\begin{equation}\label{wolf}
 R = k \, (10 g + f),\,
\end{equation}
where $g$ is the number of observed sunspot groups, $f$ the number of spots and $k$ is an observatory-related correction factor (the details depending on the actual seeing conditions, the instrument used, and the observer).

\begin{figure}
\centering
 \resizebox{8.5cm}{!}{\includegraphics{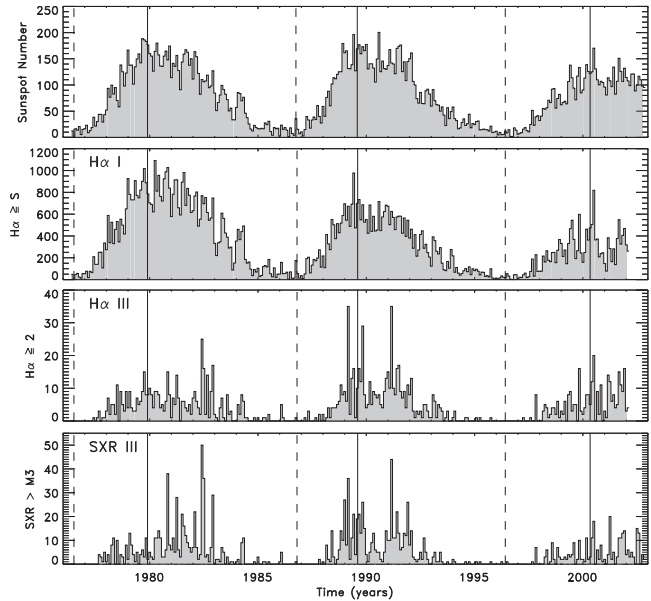}}
  \caption{Top to bottom: Monthly mean Sunspot Numbers (taken from SIDC), H$\alpha$ flares, H$\alpha$ flares of importance class $\geq$ 2, SXR flares of class $>$M3, from 1976–-2002. Solid (dashed) lines indicate the solar cycle maxima (minima). Adapted from Temmer et al. (2003).}
    \label{cycle}
\end{figure}

In principle, many other activity indices related to photospheric, chromospheric or coronal activity phenomena,
closely follow the course of the sunspot cycle, since all these activity processes are related to the underlying magnetic field and its changes over the solar cycle. In general, the magnetic field is known to be more complex during the maximum and decline phase than during the minimum and rising phase of the solar cycle (Cliver et al. 1996). The  Sunspot Number may serve as a proxy measurement of the total magnetic flux that is present at the solar surface at a given time (de Toma et al. 2000).

\subsection{Relation of Sunspot Number to other activity indices}

The variation between solar maximum and minimum varies for both monthly sunspot number and monthly averaged solar flare rate (H$\alpha$ and SXR) by about a factor of 20 (cf.\ Aschwanden 1994). However, considering only flare events of higher energy release (e.g.\ $>$ H$\alpha$ class 2, $>$ SXR class M) the fluctuations are stronger and clearly deviate from those of the Sunspot Number (cf.\ Fig.~\ref{cycle}). This indicates a non-linear response between available magnetic energy (proxy sunspots) and the flaring rate (Aschwanden 2005).

\begin{figure}
\centering
 \resizebox{7cm}{!}{\includegraphics{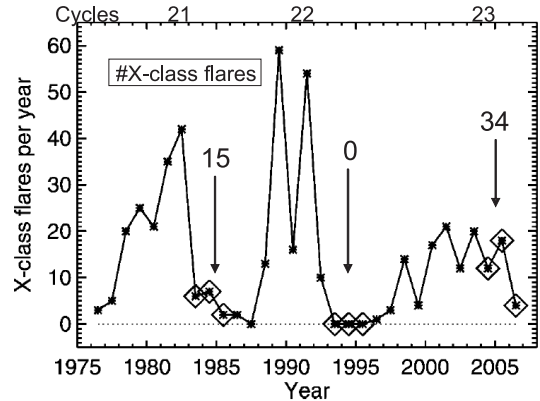}}
  \caption{Number of X-class flares for the years 1975--2007. Diamonds mark years during declining phases (1983–--1985, 1993–--1995, and 2004–--2006). A 22-yr pattern in the occurrence of the most energetic events is indicated. Adapted from Hudson (2007).
}
    \label{hudson}
\end{figure}

Interesting clues on the dynamics of the solar cycle are revealed when inspecting the peak times of several activity indices compared to the relative sunspot number. During solar cycle~21, the soft X-ray (SXR) flare occurrence as well as the SXR background flux (representing the slowly evolving flux level independent of strong flare effects) were observed to be significantly delayed with regard to the Sunspot Numbers, and a lag of 2--3~years between the peak times was revealed (e.g.\ Wagner 1988, Pearce et al.\ 1992, Aschwanden 1994, Bromund et al.\ 1995). This delay has been interpreted in terms of complexity of the coronal magnetic field in the decay phase of the solar cycle (Aschwanden 1994). However, Wilson (1993) studied SXR flare occurrences for solar cycle~22 up to the year 1992 and did not find evidence for such a delay. Veronig et al.\ (2002) investigated the soft X-ray background flux based on GOES 1-–8~\AA~measurements for the period 1975–-2003 and found only for cycle 23 a significant delay ($\sim$2 years) of the peak of the X-ray background flux with regard to the Sunspot Numbers. Likewise, shifts in the range of 10--15 months between the evolution of monthly Sunspot Number, H$\alpha$, and SXR flare rate are reported for solar cycles 19, 21, and 23 with a more prominent lag for high energetic flare events (Temmer et al.\ 2003).

Similarly, the occurrence rate of high energetic hard X-ray flare events shows a lag for cycle 21 (Bai 1993, Bromund et al.\ 1995). In a recent study, Hudson (2007) found for cycle 21 and 23 that most X-class flares are produced \textit{after} the cycle maximum (cf.\ Fig.~\ref{hudson}). From this he concluded that due to the complexity of the processes in the solar interior, which finally produce high energetic solar events, a forecasting on longer timescales is not possible. Also \v{S}vestka (1995) reported that most outstanding activity events occur after maximum, and are hence not well correlated with the peak of the solar activity cycle.


The 22-yr variation reflected in this behavior suggests a close connection to the solar dynamo and might be caused by a relic solar field (a hypothesis first proposed by Cowling 1945). Similar conclusions are drawn for the empirical Gnevyshev-Ohl rule (Cliver et al.\ 1996, Cliver \& Ling 2001, Mursula et al.\ 2001, Litvinenko \& Wheatland 2004). Possibly a 22-yr variation in the energy supply rate should be taken into account as proposed by Wheatland \& Litvinenko (2001). The question arises if the found 22-yr variation is due to the dynamic energy build-up in the corona or due to solar interior dynamics, i.e.\ related to the solar dynamo itself.

\subsection{Flare frequency distribution}

Assuming an energy storage model, a correlation between the elapsed time spanning two flare events and the size of the event is expected, whereas for a selforganized system such a correlation is not expected (Lu \& Hamilton 1991). From observations no correlation is found between successive flare events (Biesecker 1994; Crosby et al. 1998; Hudson et al. 1998; Wheatland 2000). The avalanche model of solar flares relates the power-law distributions to the scale-invariant properties of a selforganized system in a critical state (Lu \& Hamilton 1991). Since the avalanche distribution may be largely insensitive to the underlying physics of the turbulent system, the observed power-law distributions are assumed not to change over the solar cycle (see, e.g., Lu et al. 1993).

\begin{figure}
\centering
 \resizebox{11cm}{!}{\includegraphics{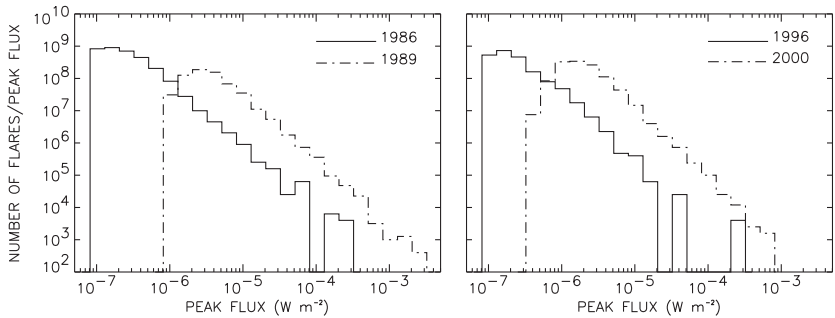}}
  \caption{The left panel shows the power-law distribution of SXR flares for solar cycle 22 (calculated from the years 1986 and 1989), the right panel for cycle 23 (1996/2000). Adapted from Veronig et al. (2002).
}
    \label{sxr-freq}
\end{figure}

For hard X-ray flares during a solar cycle an invariance of their power-law distribution as function of their energy is reported (Dennis 1985, Crosby et al.\ 1993, Lu et al.\ 1993). Likewise, for soft X-ray flares the power-law index (slope of the distribution) was found not to remarkably change between cycle 22 and 23 (Veronig et al.\ 2002). Figure~\ref{sxr-freq} shows that during times of minimum activity the power-law behavior extends to smaller peak fluxes, since during these periods also less intense flares can be detected, whereas, during maximum it extends to larger events. However, this does not influence the slope of the distribution. Christe et al.\ (2008) obtained a similar result studying the frequency distribution of RHESSI flares in the energy range of 6--12keV (thermal emission) for the years 2002--2006. The power-law index does vary between the different indices but it does not change significantly over the course of the solar cycle. From such results we may infer that solar interior dynamics dominates the occurrence frequency for the most energetic events, rather than the coronal development (see also Hudson 2007).

\subsection{Coronal mass ejections}

Coronal mass ejections (CMEs) and flares are closely related activity phenomena (e.g.\ Zhang et al.\ 2001, Temmer et al.\ 2008). Thus, it is not surprising that their occurrence rate and properties vary as well during the solar cycle. For cycle 23 (1996--2002) the CME rate increases from solar minimum (0.5/day) to maximum (6/day) by a factor of 10 and their mean and median speeds by a factor of 2 (Gopalswamy et al.\ 2003). Robbrecht, Berghmans, van der Linden (2009) found that for the same number of sunspots, more CMEs are produced especially during the decaying phase of the solar cycle.

The CME activity peak of cycle~23 shows a significant delay with respect to the sunspot cycle. Lag times varying from six months to one year are reported (Robbrecht, Berghmans, van der Linden 2009). The CME rate during cycle~23 thus tracks the solar activity cycle in amplitude but phase shifted. Since this effect was not clearly present in the activity rates of cycles~21 and 22 (Webb \& Howard 1994), it might be a possible peculiarity to cycle~23.

\begin{figure}
\centering
 \resizebox{9cm}{!}{\includegraphics{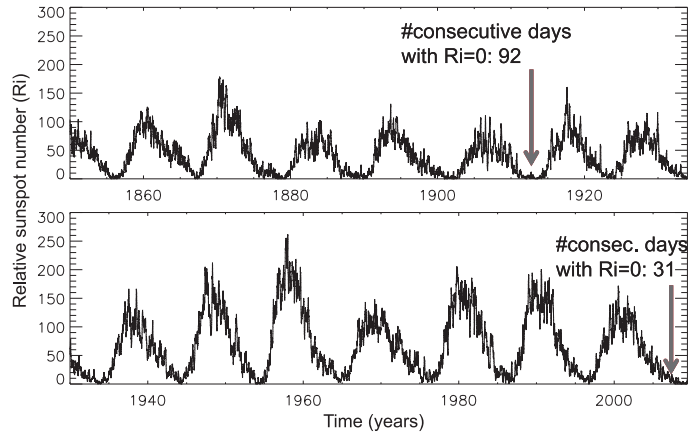}}
  \caption{Monthly smoothed Sunspot Number (taken from SIDC) for solar cycles 9--23. Number of consecutive days without visible sunspots are given for the minimum of cycle 14 (longest minimum since records) and as comparison for the minimum of cycle 23.
}
    \label{cycle-conse}
\end{figure}

\section{North-South Asymmetry}

The existence of a North-South (N-S) asymmetry is derived for various kinds of solar activity indices, like sunspots (Sunspot Numbers, areas, groups, magnetic classes, etc.), flares, prominences, and CMEs. Many studies addressed the asymmetric behavior of the solar cycle over the past 30 years, e.g.\, Garcia 1990, Joshi 1995, Atac \& \"Ozguc 2001, Temmer et al.\ 2001, 2006, Brajsa et al.\ 2005, Knaack et al.\ 2005, Carbonell 2007). From this we can safely say that the N-S asymmetry is a real effect and not a phenomenon due to random fluctuations in solar activity as was already pointed out by Carbonell et al. (1993).

As the cycle progresses, in general the northern hemisphere dominates soon after minimum and the south at the end of the cycle (e.g.\ Garcia 1990). Not only the evolution but also the average rotational periods differ between northern and southern hemisphere (e.g.\ Temmer et al.\ 2002, Joshi et al.\ 2009). The difference between the N-S activity changes from cycle to cycle with northern activity being greater during even cycles. Again this points to a 22-yr periodicity of the N-S asymmetry in solar activity, suggesting a relation to the 22-yr solar magnetic cycle (Swinson et al.\ 1986). There are hints of a long-term variation of N-S asymmetry, overlaid to the 22-yr variation, which might be of 8 solar cycles, however, to prove this longer time series are needed (Vizoso \& Ballester 1990, Li et al.\ 2009).

\section{How peculiar is the present solar minimum?}

Usoskin et al.\ (2007) pointed out that grand minima/maxima which were present during the long-term activity of the Sun are not a result of cyclic variations but defined by stochastic processes. Solanki et al.\ (2004) concluded from studies of radiocarbon concentrations, that solar activity during the past 70 years is exceptional high (the last equally high activity was more than 8,000 years ago). Hence, we are currently in the Modern Maximum (starting with 1920). However, this modern maximum is expected to end within the next 2--3 solar cycles (Abreu et al. 2008).

In Fig.~\ref{cycle-conse} we showv the monthly smoothed Sunspot Numbers (taken from SIDC) for cycles 9--23. In addition, we calculated the number of consecutive spotless days and derived a value of 92 for the minimum of cycle 14 (longest stretch for the investigated time range). During the minimum phase of cycle 23 only 31 consecutive days without sunspots are observed.

\begin{figure}
\resizebox{9cm}{!}{\includegraphics{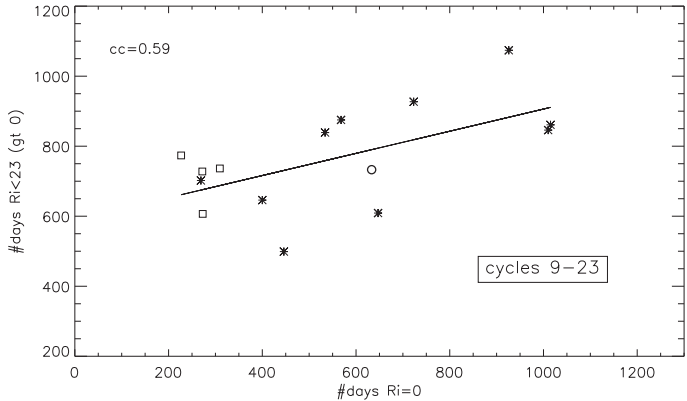}}
 \caption{Number of days without sunspots against days with Sunspot Number $<$23 (but $>$0) for the minimum phases of cycles 9--23. Cycles 19--22 are indicated as rectangles, cycle 23 as circle.}
 \label{cycle-zero}
\end{figure}

In Fig.~\ref{cycle-zero} we plot for the minimum phases of solar cycles 9--23 the number of days of zero Sunspot Number versus the days with a Sunspot Number less than 23 but $>$0 (Sunspot Number source: SIDC). A linear relation between these parameters is found with a correlation coefficient $cc$=0.59. Cycles 19--22 represent the characteristics of high minima, i.e.\ low number of days without sunspots. Compared to other cycles, cycle 23 is just average and shows no exceptional behavior during its minimum phase.

\section{Conclusions}

The occurrence of flares as well as the X-ray background flux do not evolve one-to-one in phase with the relative sunspot number. A 22-year variation is derived for the occurrence of high energetic events as well as for the N-S asymmetry behavior, indicating a strong relation to the magnetic activity cycle. At the present for CMEs this variation is not clearly derived and more cycles need to be studied. Solar interior processes or dynamo related processes might drive such a 22-year variation (cf.\ Hudson 2007), which gives important constraints to solar dynamo models.

At present there is, in addition to ground-based observatories, a fleet of spacecrafts dedicated to observe the Sun: SOHO, GOES, RHESSI, ACE, TRACE, STEREO, Hinode, Coronas-F, Proba2, etc. Space era started during times of high minima, so the current prolonged minimum disappoints from the point of view of observational possibilities that are currently available. However, there is nothing peculiar about the minimum of cycle 23 when compared with previous cycles. But how about the next cycle(s), are we at the end of high activity as proposed by Abreu et al. (2008)?

\acknowledgements M.T. is a recipient of an APART-fellowship of the Austrian Academy of Sciences at the Institute of Physics, University of Graz (APART 11262).

\bibliographystyle{apj}

\end{document}